\documentstyle[prd,aps,preprint,epsfig,amssymb]{revtex}
\begin{document}
%%%%%%%%%%%%%%%%%%%%%%%%%%%%%%%%%%%%%%%%%%%%%%%%%%%%%%%%
\newcommand{\TeV}{\,{\rm TeV}}
\newcommand{\GeV}{\,{\rm GeV}}
\newcommand{\MeV}{\,{\rm MeV}}
\newcommand{\keV}{\,{\rm keV}}
\newcommand{\eV}{\,{\rm eV}}
\def\ap{\approx}
\def\bea{\begin{eqnarray}}
\def\eea{\end{eqnarray}}
\def\bec{\begin{center}}
\def\ec{\end{center}}

%%%%%%%%%%%%%%%%%%%%%%%%%%%%%%%%%%%%%%%%%%%%%%%%%%%%%%%%
\def\f#1#2{\frac{#1}{#2}}
\def\p{\partial}
\def\l{\left}
\def\r{\right}
\def\a{\alpha}
\def\Lm{\Lambda}
\def\tr{{\rm Tr}}
\def\Order#1{{\cal O}\l(#1\r)}
\def\t{\tilde}
\def\Tilde#1{\stackrel{\sim}{#1}}
\def\Infi#1{\stackrel{\infty}{#1}}
\def\T{{\mathcal{T}}}
%%%%%%%%%%%%%%%%%%%%%%%%%%%%%%%%%%%%%%%%%%%%%%%%%%%%%%%%

\def\pC{\tilde{\chi}^+}
\def\nC{\tilde{\chi}^-}
\def\pnC{\tilde{\chi}^{\pm}}
\def\Ne{\tilde{\chi}^0}
\def\snu{\tilde{\nu}}
\def\tN{\tilde N}
\def\ler{\lesssim}
\def\gtr{\gtrsim}
\def\beq{\begin{equation}}
\def\eeq{\end{equation}}
\def\haf{\frac{1}{2}}
\def\plb#1#2#3#4{#1, Phys. Lett. {\bf #2B} (#4) #3}
\def\plbb#1#2#3#4{#1 Phys. Lett. {\bf #2B} (#4) #3}
\def\npb#1#2#3#4{#1, Nucl. Phys. {\bf B#2} (#4) #3}
\def\prd#1#2#3#4{#1, Phys. Rev. {\bf D#2} (#4) #3}
\def\prl#1#2#3#4{#1, Phys. Rev. Lett. {\bf #2} (#4) #3}
\def\mpl#1#2#3#4{#1, Mod. Phys. Lett. {\bf A#2} (#4) #3}
\def\rep#1#2#3#4{#1, Phys. Rep. {\bf #2} (#4) #3}
\def\lpp{\lambda''}
\def\ccg{\cal G}
\def\slash#1{#1\!\!\!\!\!/}
\def\rpv{\slash{R_p}}
\def\ler{\lesssim}
\def\gtr{\gtrsim}
\def\beq{\begin{equation}}
\def\eeq{\end{equation}}
\def\haf{\frac{1}{2}}
\def\plb#1#2#3#4{#1, Phys. Lett. {\bf #2B} (#4) #3}
\def\plbb#1#2#3#4{#1 Phys. Lett. {\bf #2B} (#4) #3}
\def\npb#1#2#3#4{#1, Nucl. Phys. {\bf B#2} (#4) #3}
\def\prd#1#2#3#4{#1, Phys. Rev. {\bf D#2} (#4) #3}
\def\prl#1#2#3#4{#1, Phys. Rev. Lett. {\bf #2} (#4) #3}
\def\mpl#1#2#3#4{#1, Mod. Phys. Lett. {\bf A#2} (#4) #3}
\def\rep#1#2#3#4{#1, Phys. Rep. {\bf #2} (#4) #3}
\def\lpp{\lambda''}
\def\ccg{\cal G}
\def\slash#1{#1\!\!\!\!\!/}
\def\rpv{\slash{R_p}}
\def\pslash{p\hspace{-2.0mm}/}
\def\qslash{q\hspace{-2.0mm}/}
\newcommand{\imag}{{\rm Im}~}
\newcommand{\real}{{\rm Re}~}

%%%%%%%%%%%%%%%%%%%%% Main Text %%%%%%%%%%%%%%%%
\setcounter{page}{1}
%\renewcommend{\arraystretch}{1.3}
\draft
%\widetext
\preprint{KAIST-TH 02/18}

\title{\large \bf One loop gauge couplings in ${\rm AdS}_5$}

\author{Kiwoon Choi\footnote{kchoi@muon.kaist.ac.kr} and
Ian-Woo Kim\footnote{iwkim@muon.kaist.ac.kr}}
\address{Department of Physics,
Korea Advanced Institute of Science and Technology\\ Daejeon
305-701, Korea}

\date{\today}
\tightenlines
\maketitle
%%%%%%%%%%%%%%%%%%%%%%%%%%%%%%%%%%%%%%%%%%%%%%%%%%%%%%%%%%%%%
\begin{abstract}
%%%%%%%%%%%%%%%%%%%%%%%%%%%%%%%%%%%%%%%%%%%%%%%%%%%%%%%%%%%%%
We calculate the full 1-loop corrections to the low energy coupling of
bulk gauge boson in a slice of ${\rm AdS}_5$
which are
induced by generic 5-dimensional scalar, Dirac fermion, and vector
fields with arbitrary $Z_2\times Z_2^{\prime}$ orbifold boundary conditions.
In supersymmetric limit, our results correctly reproduce the results
obtained by an independent method based on 4-dimensional effective supergravity.
This provides a nontrivial check of our results and assures
the regularization scheme-independence of the results.
%%%%%%%%%%%%%%%%%%%%%%%%%%%%%%%%%%%%%%%%%%%%%%%%%%%%%%%%%%%%%
\end{abstract}
%%%%%%%%%%%%%%%%%%%%%%%%%%%%%%%%%%%%%%%%%%%%%%%%%%%%%%%%%%%%%

\section{Introduction}
Models with extra dimension have  provided a new insight on
the large scale hierarchy between the Weak scale
$M_W\sim 10^2$ GeV and the Planck scale $M_{Pl}\sim 10^{18}$ GeV.
In this regard, the Randall-Sundrum model (RS1) is particularly interesting
as it explains the Weak to Planck scale ratio using the warped 5D geometry\cite{Randall:1999ee}:
\beq
\label{5dmetric}
ds^2=G_{MN}dx^Mdx^N=e^{-2kR|y|}g_{\mu\nu}dx^\mu dx^\nu+R^2 dy^2\,,
\eeq
where $-\pi \leq y \leq \pi$, $k$ is the AdS curvature and $R$ is the
orbifold radius.
In this spacetime background,
4-dimensional (4D) graviton is localized near the
UV brane at $y=0$ whose cutoff mass scale $M_{UV}$
is of order the 5D Planck scale.
On the other hand, in the original RS1 model,
all the standard model (SM) fields are assumed to be confined on
the IR brane at $y=\pi$ whose cutoff scale  $M_{IR}\sim e^{-\pi kR}M_{UV}$.
Then with a moderately large value of $kR$ ($\sim 12$), the model
can generate the large scale hierarchy
$M_{Pl}/M_W\sim M_{UV}/M_{IR}\sim 10^{16}$ without any severe fine tuning of
the fundamental parameters.

An apparent drawback of the original RS1 model is that
one has to abandon the attractive possibility that the SM
gauge couplings $g_a^2$ ($a=1,2,3$) are unified at high energy scale
through {\it the quantum corrections calculable within
the model}.
Experimental data show that
$g_a^2$ at $M_W$ differ
from each other by order unity:
\beq
\frac{1}{g_a^2(M_W)}-\frac{1}{g_b^2(M_W)}
={\cal O}(1)\, \quad (a\neq b)\,.
\eeq
On the other hand, the size of quantum corrections to $1/g_a^2$
which are calculable within the RS1 model is
\beq
\Delta \left(\frac{1}{g_a^2}\right) \,=\,{\cal O}
\left(\frac{1}{8\pi^2}\ln (M^2_{IR}/M^2_W)\right) \,=\,{\cal O}\left(
\frac{1}{8\pi^2}\right)\,,
\eeq
so the RS1 model does not give any insight on why the SM gauge couplings
at $M_W$ differ from each other by order unity.

It has been noted recently
\cite{Pomarol:2000hp,Randall:2001gb,Choi:2002wx,Goldberger:2002cz,Agashe:2002bx,Choi:2002zi}
that one can achieve the gauge unification, while still solving
the hierarchy problem, within the 5D effective field theory on
${\rm AdS}_5$
if the SM gauge bosons propagate in 5D bulk spacetime.
In such case, the size of quantum corrections
calculable within the model is
\beq
\Delta \left(\frac{1}{g_a^2}\right)\,=\,
{\cal O}\left(\frac{1}{8\pi^2}\ln (M^2_{Pl}/M^2_W)\right)
\,=\,{\cal O}(1)\,,
\eeq
as in the case of conventional 4D grand unified theories (GUT).
This allows that the observed differences of gauge couplings
are explained in terms
 of quantum corrections  which are calculable within the model.

Calculation of the 1-loop corrections to gauge coupling
in ${\rm AdS}_5$ was first attempted in \cite{Pomarol:2000hp} for a GUT model
in which all gauge-charged matter fields are confined on the UV brane.
The computation involves a Pauli-Villars regulator with regulator mass
$\Lambda_{PV}\ll k$, so could catch only the corrections at scales
significantly below $k$.
In \cite{Randall:2001gb}, a momentum cutoff
depending on the position in  5-th dimension was proposed to regulate
the 1-loop corrections.
Though intuitively sensible,
it is difficult to isolate the regulator-independent part from
the regulator-dependent total corrections in this regularization, which makes
the interpretation of the results unclear.
In \cite{Choi:2002wx,Choi:2002zi}, the 1-loop corrections have been computed
for generic supersymmetric gauge theory on ${\rm AdS}_5$ using
the gauged $U(1)_R$ symmetry and chiral anomaly in 5D supergravity (SUGRA)
and also the known properties of
gauge couplings in 4D effective SUGRA.
In this approach, one could obtain the 1-loop corrections
(including those from scales between $k$ and
the 5D cutoff scale $\Lambda>k$) in obviously regulator-independent manner.
In \cite{Goldberger:2002cz,Agashe:2002bx}, 1-loop corrections in 5D scalar QED
on ${\rm AdS}_5$ have been computed (using dimensional regularization and
also Pauli-Villars regularization)
and the results are nicely interpreted in terms of AdS/CFT correpondence.

In this paper, we present the full 1-loop corrections to the low energy
coupling of bulk gauge boson in a slice of ${\rm AdS}_5$ which are induced by
generic 5D scalar, Dirac fermion and vector fields with
arbitrary $Z_2\times Z_2^{\prime}$ orbifold boundary
condition. To be explicit, we adopt dimensional regularization\cite{GrootNibbelink:2001bx},
but the results should be independent of the used
regularization scheme as they correspond to the scheme-independent
corrections calculable within 5D effective field theory.
When applied to supersymmetric case\cite{Altendorfer:2000rr,Gherghetta:2000qt}, our results correctly reproduce the
expressions which are obtained in a completely independent
approach based on 4D effective SUGRA.
This provides a nontrivial check of our results, and
also assures the scheme-independence of the results.
We also note that the subtraction scales of log divergences
at two orbifold fixed points, i.e. $y=0$ and $\pi$,
differ by the warp factor $e^{-\pi kR}$.
This is physically expected, and can be confirmed by comparing the results
with those of Pauli-Villars regularization
as well as with the results of 4D SUGRA calculation.

The organization of this paper is as follows.
In section II, we set up the notations for 5D gauge theory
on a slice of ${\rm AdS}_5$ including the Kaluza-Klein (KK)
analysis for generic 5D scalar, Dirac fermion, and
vector fields with arbitrary $Z_2\times Z_2^{\prime}$ orbifold
boundary condition.
In section III, we present our main results, i.e. 1-loop gauge couplings
in ${\rm AdS}_5$ induced by generic 5D fields,
obtained using the background field method with dimensional regularization.
In section IV,
we consider the supersymmetric limit in order to confirm that our results
correctly reproduce the results from 4D SUGRA calculation,
and conclude in section V.

\section{Gauge theory on a slice of ${\rm AdS}_5$}

The model we study is a 5D gauge theory defined
on a slice of ${\rm AdS}_5$ with spacetime metric (\ref{5dmetric}),
containing generic gauge-charged 5D scalar, fermion and vector fields
with arbitrary $Z_2\times Z_2^{\prime}$ boundary condition.
The lagrangian is given by
\bea
\int d^4 x d y \sqrt{-G} \left[ -\f{1}{4g_{5a}^2} F^{aMN} F^a_{MN}
- \f{1}{2}D_M \phi D^M \phi - \f{1}{2}m^2_\phi \phi^2
- i \bar{\psi} (\gamma^M D_M +m_\psi) \psi \right],
\eea
where $D_M$ is the covariant derivative containing the gauge connections
as well as the spin connection of ${\rm AdS}_5$.
We parametrize the masses of scalar and fermion fields as
\bea
\label{mass}
m^2_\phi=A^2k^2 +  \f{2k}{R} \left[ \,
B_0 \delta(y) - B_\pi \delta(y-\pi) \,\right],
\quad
m_\psi &=& Ck \epsilon(y),
\eea
where $\epsilon(y) = y/|y|$,
$B_0$ and $B_\pi$ are the brane mass parameters at $y=0$ and $y=\pi$,
respectively, and $c$ is the fermion kink mass parameter.
The 5D fields in the model can have arbitrary
$Z_2\times Z_2^{\prime}$ orbifold boundary condition,
\bea
&&\phi(-y)=Z_\phi \phi(y)\,,\quad
\phi(-y')=Z^{\prime}_\phi \phi(y')\,,
\nonumber \\
&&\psi(-y)=Z_\psi \gamma_5\psi(y)\,,\quad
\psi(-y')=Z^{\prime}_\psi \gamma_5\psi(y')\,,
\nonumber  \\
&& A^a_\mu(-y)=Z_a A^a_\mu(y)\,,\quad
A^a_\mu(-y')=Z^{\prime}_a A^a_\mu(y')\,,
\eea
with $Z_\Phi=\pm 1$ and $Z^{\prime}_\Phi=\pm 1$ for
$\Phi=\{\,\phi,\psi, A^a_M\,\}$ and $y'=y-\pi$.
Though we are interested
in the low energy coupling of $A^a_\mu$ having
$Z_a=Z^{\prime}_a=1$, there can be 5D vector fields having
other $Z_2\times Z_2^{\prime}$ parity which are charged for
the gauge fields with $Z_a=Z^{\prime}_a=1$.
Note that the brane mass of scalar field at $y=0$ ($y=\pi$)
is relevant only when $Z_\phi=1$ ($Z^\prime_\phi=1$).

The KK specturm of bulk fields on a slice of ${\rm AdS}_5$
has been discussed in detail
in \cite{Gherghetta:2000qt}.
It is rather straightforward to generalize the analysis of
\cite{Gherghetta:2000qt} to the field with arbitrary $Z_2\times Z_2^{\prime}$ parity.
A generic 5D field $\Phi$ can be decomposed as
$$\Phi( x, y ) = \sum \Phi_n(x) f_n (y),$$
where the KK wavefunction $f_n$ satisfies
\beq
\l[\, - e^{s k R|y| } \p_y \l(e^{-s k R|y| } \p_y \r) +
R^2k^2\hat{M}^2_\Phi \,\r] f_n
= R^2e^{2 k R |y| } m_n^2 f_n
\eeq
for the KK mass eigenvalue $m_n$.
Here
\bea
s=\{2,4,1,1\}
\eea
and the bulk mass parameters
\bea
\hat{M}^2_\Phi
= \{0 , A^2, C(C+1), C(C-1) \}
\eea
for
\bea
\Phi =
\{ A_\mu , \phi , e^{-2 k R |y|} \psi_{L}, e^{-2kR|y|}\psi_R\}\
\quad
(\psi_{L,R}=\frac{1}{2}(1\pm\gamma_5)\psi).
\nonumber
\eea
This determines $f_n$ to be
\beq
f_n (y) = e^{s k R |y| /2 } \l[ J_\a\l( \f{m_n}{k} e^{ k R |y| } \r) +
b_\a\l( m_n \r) Y_\a\l( \f{m_n}{k} e^{k R |y| } \r) \r]\,,
\label{wavefn}
\eeq
where
\beq
\a = \sqrt{(s/2)^2 + \hat{M}^2_\Phi}. \label{al}
\eeq
To determine the corresponding  KK mass spectrum,
one needs to impose the orbifold boundary condition.
Parity-even condition under the reflection at
$y = 0$ or $\pi$ leads to
\beq
\label{evenbc}
\f{d f_n}{d y } =r k R f_n \quad {\rm at} \,\, y=0 \,\,\, {\rm or} \,\,\,
 \pi\,,
\eeq
where
\bea
r = \{\, 0 , B_0~\,{\rm or}~\,B_\pi, -C, C \,\}
\eea
for
$$\Phi =\l\{ A_\mu , \phi , e^{-2 k R |y|} \psi_{L}, e^{-2kR|y|}\psi_R\r\}.$$
Then using Eqs. (\ref{wavefn}) and (\ref{evenbc}),
one finds
\beq
b_{\a}(m_n) = - \f{ (\f{s}{2} -r ) J_\a \l( \f{m_n}{k} e^{k R \tilde{y}} \r)
+ \f{m_n}{k} e^{k R \tilde{y}} J'_\a \l( \f{m_n}{k} e^{k R \tilde{y}} \r) }
{(\f{s}{2} -r ) Y_\a \l( \f{m_n}{k} e^{k R \tilde{y}} \r)
+ \f{m_n}{k} e^{k R \tilde{y}} Y'_\a \l( \f{m_n}{k} e^{k R \tilde{y}} \r)}\,,
\eeq
where $\tilde{y}=0$ or $\pi$.
Parity-odd condition under the reflection at $y= 0$ or $\pi$ leads to
\beq
f_n = 0 \quad {\rm at}\,\,  y=0\,\, {\rm or} \,\, \pi\,,
\eeq
yielding
\beq
b_\a(m_n) = - \f{J_\a \l( \f{m_n}{k} e^{k R \tilde{y}} \r)}{Y_\a \l( \f{m_n}{k}
e^{k R \tilde{y}} \r)}\,.
\eeq

With the above results, the KK spectrum of 5D field $\Phi$ can be determined
by the so-called $N$-function $N(q)=N(-q)$ which has simple zeros
at $q=\pm m_n\neq 0$:
\bea
\label{nzero}
N(m_n)=0\,.
\eea
If there exists a massless
mode, $N$ has a double zero at $q=0$.
%\beq
% \f{ (\f{s}{2} -r_0 ) J_\a \l( \f{m_n}{k} \r)
%+ \f{m_n}{k} J'_\a \l( \f{m_n}{k}\r) }
%{(\f{s}{2} -r_0 ) Y_\a \l( \f{m_n}{k}  \r)
%+ \f{m_n}{k} Y'_\a \l( \f{m_n}{k} \r)}=
% \f{ (\f{s}{2} -r_\pi ) J_\a \l( \f{m_n}{k} e^{\pi k R} \r)
%+ \f{m_n}{k} e^{\pi k R} J'_\a \l( \f{m_n}{k} e^{\pi k R} \r) }
%{(\f{s}{2} -r_\pi ) Y_\a \l( \f{m_n}{k} e^{\pi k R} \r)
%+ \f{m_n}{k} e^{\pi k R} Y'_\a \l( \f{m_n}{k} e^{\pi k R} \r)},
%\eeq
For later use, here we summarize the $N$-functions for
all $Z_2\times Z_2^{\prime}$ boundary conditions
of the corresponding 5D field.
Let $r_0$ and $r_\pi$ denote the mass parameters at $y=0$ and
$\pi$, respectively, given by
\bea
\label{r}
r_0=\{0,B_0,-C,C\},\quad
r_\pi=\{0,B_\pi,-C,C\}
\eea
for
$$\Phi =\l\{ A_\mu , \phi , e^{-2 k R |y|} \psi_{L}, e^{-2kR|y|}\psi_R\r\}.$$
%If $r_0=r_\pi$ and $\a = |s/2 - r|$, $\Phi$ has a  massless mode and
%its $N$ function is given by
%\bea
%N_{\Phi^{(++)}}(q) =
%J_{|s/2 -r-1|}\l( \f{q}{k} \r)Y_{|s/2-r-1|}\l( \f{q}{ke^{-\pi kR}} \r)
%- J_{|s/2-r-1|}\l( \f{q}{ke^{-\pi kR}} \r)Y_{|s/2-r-1|} \l( \f{q}{k} \r)\,,
%\eea
%while in other cases
The $N$-function for  $(Z_\Phi,Z^{\prime}_\Phi)=(+,+)$ is given by
\bea
N_{++} (q) &=&
-\l\{(\frac{s}{2}-r_0)J_\alpha\l(\f{q}{k} \r)+\f{q}{k} J'_\alpha\l(\f{q}{k}\r) \r\}
\l\{(\frac{s}{2}-r_\pi)Y_\alpha\l(\f{q}{T} \r)+
\f{q}{T} Y'_\alpha\l(\f{q}{T}\r) \r\}
\nonumber \\
&&+\l\{(\frac{s}{2}-r_\pi)J_\alpha\l(\f{q}{T} \r)+
\f{q}{T} J'_\alpha\l(\f{q}{T}\r) \r\}
\l\{(\frac{s}{2}-r_0)Y_\alpha\l(\f{q}{k} \r)+\f{q}{k} Y'_\alpha\l(\f{q}{k}\r) \r\}
\label{++bc}
\eea
where $T=ke^{-\pi kR}$.
As for the fields with other boundary conditions, i.e.
$(Z_\Phi,Z^{\prime}_\Phi)=(+,-),(-,+),(-,-)$, we find
\bea
N_{+-} (q) &=&  -Y_\alpha\l(\f{q}{T}\r)
\l[\,(\frac{s}{2}-r_0)
J_\alpha\l(\f{q}{k} \r)+\f{q}{k} J^{\prime}_\alpha\l(\f{q}{k}\r)\, \r]
\nonumber \\
&&+J_\alpha\l(\f{q}{T}\r)
\l[\,(\frac{s}{2}-r_0)Y_\alpha\l(\f{q}{k} \r)+
\f{q}{k} Y^{\prime}_\alpha\l(\f{q}{k}\r)\, \r]\,,
\nonumber \\
N_{-+} (q) &=&
J_\alpha\l(\f{q}{k}\r)
\l[\,(\frac{s}{2}-r_\pi)Y_\alpha\l(\f{q}{T} \r)+
\f{q}{T} Y^{\prime}_\alpha\l(\f{q}{T}\r)\, \r]
\nonumber \\
&&-Y_\a\l(\f{q}{k}\r)
\l[\,(\frac{s}{2}-r_\pi)
J_\alpha\l(\f{q}{T} \r)+\f{q}{T}
J^{\prime}_\alpha\l(\f{q}{T}\r)\,\r]\,,
\nonumber \\
N_{--} (q) &=& J_{\alpha} \l( \f{q}{k} \r)
  Y_{\alpha} \l( \f{q}{T} \r)
-   J_{\alpha} \l( \f{q}{T} \r)
  Y_{\alpha} \l( \f{q}{k} \r).
\label{otherbc}
\eea

As we will see in the next section, one can choose an appropriate gauge fixing
to make that the KK spectrum of $A_5$ is determined by the $N$-function
of 5D scalar field $\phi$ with a specific mass:
\bea
\label{same}
N_{A_5}=N_\phi \quad{\rm for} \,\,\, m_\phi^2=-4k^2+\f{4k}{R}(\delta(y)-\delta(y-\pi)).
\eea
In fact, one needs to know the asymtotic bahaviors of these $N$-functions
at $|q|\rightarrow\infty$
to regulate the UV divergence
and also the behaviors at $|q|\rightarrow 0$ to find the 1-loop couplings
in the IR limit.
Some properties of the $N$-functions
including those asymtotic behaviors are summarized in
Appendix A.

\section{one loop effective couplings}

In this section, we calculate the 1-loop effective coupling of
gauge field zero mode in ${\rm AdS}_5$ using the
background field method\cite{Peskin:ev} with dimensional regularization\cite{GrootNibbelink:2001bx}.
Let us first describe the calculation scheme.
We split the gauge field as
\bea
A_M^a=\bar{A}^a_M+\tilde{A}_M^a\,,
\eea
where $\bar{A}_M^a$ denotes the background gauge field in the gauge
$\bar{A}^a_5=0$ and
$\tilde{A}_M^a$ is the quantum fluctuation.
We choose the gauge fixing term
\bea
-\frac{1}{2g^2_{5a}}
\int d^5x \sqrt{-G}\,\l[\,e^{2kR|y|}g^{\mu\nu}D_\mu\tilde{A}^a_\nu
+\f{e^{2kR|y|}}{R^2}\p_y(e^{-2kR|y|}\tilde{A}_5)\,\r]^2
\eea
where $D_\mu$ is defined by the background gauge field $\bar{A}^a_\mu$.
The corresponding ghost action is given by
\bea
\int d^5x\sqrt{-G}\l[\, e^{2kR|y|}\bar{\xi}^aD^2\xi^a+\f{e^{2kR|y|}}{R^2}\bar{\xi}^a
\p_y(e^{-2kR|y|}\p_y\xi^a)\,\r]\,,
\eea
where $D^2=g^{\mu\nu}D_\mu D_\nu$.
It is then straightforward to find the following gauge-fixed action
which are quadratic in $\tilde{A}_\mu^a,\t{A}^a_5$ and $\xi^a$:
\bea
\int d^5x\,&&\l[
-\frac{1}{4g^2_{5a}}
\,\l(\, -2R\tilde{A}^a_\mu D^2\tilde{A}^{a\mu}
+4Rf_{abc}\bar{F}^a_{\mu\nu}\tilde{A}^{b\mu}\tilde{A}^{c\nu}
-\f{2}{R}\t{A}^a_\mu\p_y(e^{-2kR|y|}\p_y)\t{A}^{a\mu}\r.\r.
\nonumber \\
&&\quad \l.\l.-\f{2}{R}e^{-2kR|y|}\t{A}^a_5D^2\t{A}^a_5-
\f{2}{R^3}e^{-2kR|y|}\t{A}^a_5\p_y^2(e^{-2kR|y|}\t{A}^a_5)\,\r)\r.
\nonumber \\
&&\l.\quad+e^{-2kR|y|}R\l\{\bar{\xi}^aD^2\xi^a-
\f{1}{R^2}\bar{\xi}^a\p_y(e^{-2kR|y|}\p_y\xi^a)\r\}
\,\r]
\eea
The action of scalar and fermion fields can be written as
\bea
\int d^5x \,\,&&\l[\,e^{-2kR|y|}R\f{1}{2}\phi
(D^2+\f{1}{R^2}e^{2kR|y|}\p_ye^{-4kR|y|}\p_y-e^{-2kR|y|}m^2_\phi)\phi\r.
\nonumber \\
&&-e^{-3kR|y|}R(\bar{\psi}_Li\gamma^\mu D_\mu\psi_L+
\bar{\psi}_Ri\gamma^\mu D_\mu\psi_R)
-e^{-4kR|y|}(\bar{\psi}_Li\gamma^5\p_y\psi_R
+\bar{\psi}_Ri\gamma^5\p_y\psi_L)
\nonumber \\
&&\l.-iRe^{-4kR|y|}m_\psi(\bar{\psi}_L\psi_R+\bar{\psi}_R\psi_L)\,\r]
\eea
Note that the quadratic action of $\t{A}_5^a$ has the same form
as the action of 5D real scalar $\phi$ with
$m_\phi^2=-4k^2+4kR^{-1}(\delta(y)-\delta(y-\pi))$,
justifying the relation (\ref{same}).

One-loop effective action of the gauge field zero mode
can be obtained by integrating out all quantum fluctuation
fields at 1-loop order.
This procedure yields
\bea
S_{eff}=\int d^4x \l(\,-\f{\pi R}{4g^2_{5a}}{F}^{a\mu\nu}{F}^a_{\mu\nu}\,\r)+
\Gamma_\phi [ {A}_\mu ]
+\Gamma_\psi [ {A}_\mu ] +\Gamma_A [ {A}_\mu ],
\eea
where the first term is obviously the tree level action, and
$\Gamma_\phi$, $\Gamma_\psi$ and $\Gamma_A$ represent
the 1-loop corrections due to the loops of
$\phi$, $\psi$, and $A^a_M$ (and also
the ghost fields $\xi^a$, $\bar{\xi}^a$), respectively:
\bea
\label{1loopcontribution}
i\Gamma_\phi &=& -\f{1}{2} \tr_\phi \ln \l( -D^2 +M^2(\phi) \r) \,,\nonumber\\
i\Gamma_\psi &=&  \f{1}{2} \tr_\psi \ln \l( -D^2 +M^2(\psi) + {F}_{\mu\nu}
J_{1/2}^{\mu\nu} \r)\,, \nonumber\\
i\Gamma_A &=&
-\f{1}{2} \tr_{A^\mu} \ln \l ( -D^2 + M^2(A_\mu) +
{F}_{\mu\nu}J_1^{\mu\nu}\r)\,,
\nonumber \\
&&-\f{1}{2} \tr_{A_5} \ln \l( -D^2 + M^2(A_5) \r)
+ \tr_{\xi,\bar{\xi}} \ln \l( -D^2 + M^2(\xi)  \r)\,.
\eea
Here we replace the background gauge field $\bar{A}_\mu^a$ by
un-barred $A^a_\mu$, and $M^2(\Phi)$ is the mass-square operator
whose eigenvalues $m_n^2$ are determined by the zeros of
the corresponding $N$-function.
$J^{\mu\nu}_{j}$ is the 4D Lorentz spin generator
normalized as ${\rm tr}(J_j^{\mu\nu}J_j^{\rho\sigma})=
C(j) (g^{\mu\rho}g^{\nu\sigma}-g^{\mu\sigma}g^{\nu\rho})$
where $C(j)=(0,1,2)$ for $(j=0,1/2,1)$.

The above 1-loop effective action is divergent, so need to be regulated.
As in the case of flat 5D orbifold,
the UV divergence structure of 5D gauge theory on ${\rm AdS}_5$ is given by
\bea
\label{divergence}
-\int d^5x \sqrt{-G} \,\l[\,\f{\gamma_a}{96\pi^3}\Lambda F^{a}_{MN}F^{aMN}
+\f{\ln\Lm}{32\pi^2}\l(\lambda_0\f{\delta(y)}{\sqrt{G_{55}}}
+\lambda_\pi\f{\delta(y-\pi)}{\sqrt{G_{55}}}\r)
 F^a_{\mu\nu}F^{a\mu\nu}\,\r]
\eea
where the coefficient of linear divergence ($\gamma_a$) is highly sensitive
to the used regularization scheme, while those of log divergences
at fixed points ($\lambda_{0,\pi}$) are scheme-independent.
In dimensional regularization, $\gamma_a=0$,
however this does not have any special physical meaning.
As for the coefficients of log divergences, it is straightforward to find\cite{Contino:2001si}
\bea
\label{logdivergence}
\lambda_0=&&\frac{1}{24}\l[T_a(\phi_{++})
+T_a(\phi_{+-})-T_a(\phi_{-+})-T_a(\phi_{--})
\r] \nonumber \\
&&-\f{23}{24}\l[T_a(A^M_{++})+T_a(A^M_{+-})-T_a(A^M_{-+})-T_a(A^M_{--})\r]
\,,\nonumber \\
\lambda_\pi=&&\f{1}{24}\l[T_a(\phi_{++})-T_a(\phi_{+-})+
T_a(\phi_{-+})-T_a(\phi_{--})\r]\nonumber \\
&&-\f{23}{24}\l[T_a(A^M_{++})-T_a(A^M_{+-})+T_a(A^M_{-+})-T_a(A^M_{--})\r]\,,
\eea
where $T_a(\Phi)={\rm Tr}(T_a^2)$ for the gauge group representation
given by $\Phi$,
 $\phi_{zz^{\prime}}$  ($z,z^{\prime}=\pm$)
is 5D {\it real} scalar field with $Z_2\times Z_2^{\prime}$ parity
$(z,z^{\prime})$, and $A^M_{zz^{\prime}}$ is 5D real vector field.

With the UV divergences given by (\ref{divergence}),
the low energy effective gauge coupling can be written
as
\bea
\f{1}{g_a^2(p)}=&&\l[\,\f{1}{g^2_{5a}(\Lambda)}+
\f{\gamma_a\Lambda}{24\pi^3}\,\r]\pi R
+\f{1}{g_{0a}^2(\Lambda)}+\f{1}{g_{\pi a}^2(\Lambda)}
%+\,\f{\lambda_0+\lambda_\pi}{8\pi^2}\ln\Lambda\,\r]
\nonumber \\
&&+\f{1}{8\pi^2}\bar{\Delta}_a(p,A,B_0,B_\pi,C,k,R,\ln\Lambda)+{\cal O}(1/\Lambda)
\eea
where
$p$ is the 4D momentum of the external gauge
boson zero mode, $g^2_{0a}(\Lm)$ and $g^2_{\pi a}(\Lm)$
denote the bare brane gauge couplings at
the orbifold fixed points $y=0$ and $y=\pi$, respectively,
and ${\cal O}(1/\Lm)$  stands for the part suppressed by $1/\Lambda$.
Here the log-divergent piece of (\ref{divergence}) and also
the conventional momentum running and finite KK threshold corrections
are all encoded in $\bar{\Delta}_a$.
The bare brane couplings $g_{0a}^2(\Lambda)$ and $g_{\pi a}^2(\Lambda)$
can be interpreted as the Wilsonian brane couplings at $\Lambda$
in the metric frame of $G_{MN}$ (see Eq. (\ref{5dmetric})).
However, when measured in the metric frame of 4D massless graviton
$g_{\mu\nu}=e^{2kR|y|}G_{\mu\nu}$, they should be
interpreted as the Wilsonian couplings at different scales,
$g_{0a}^2$ at the scale $\Lambda$ and $g_{\pi a}^2$ at
the rescaled scale $e^{-\pi kR}\Lambda$.
One can then assume that $g_{0a}^2$ and $g_{\pi a}^2$
are of order $8\pi^2$ \cite{Chacko:1999hg}, so
\bea
\f{1}{g_a^2(p)}\,=\,
\f{\pi R}{\hat{g}_{5a}^2}+
\f{1}{8\pi^2}\bar{\Delta}_a(p,A,B_0,B_\pi,C,k,R,\ln(\Lambda))+{\cal O}
\l(\f{1}{8\pi^2}\r)\,,
\eea
where
$$
\f{1}{\hat{g}_{5a}^2}=\f{1}{g^2_{5a}}+\f{\gamma_a\Lm}{24\pi^3}
$$
are the bare bulk couplings which are {\it not} calculable within
5D effective field theory.
In the low momentum limit $p\ll m_{KK}$ where $m_{KK}$ is the
KK threshold scale which corresponds to the \,{\it lowest nonzero}\,
KK mass, the calculable one-loop correction $\bar{\Delta}_a$
can be written as
\bea
\label{calculablepiece}
&&\bar{\Delta}_a(p,A,B_0,B_\pi,C,k,R,\ln\Lambda)
\nonumber \\
%&=&
%\tilde{\Delta}_a(
%A,B_0,B_\pi,k,R)+(\lambda_0+\lambda_\pi)\ln\Lambda-b_a\ln p+
%{\cal O}\l(\f{p^2}{m_{KK}^2}\r)
%\nonumber \\
&=&\Delta_a(A,B_0,B_\pi,C,k,R,\ln\Lambda)+b_a\ln\l({\Lambda}/{p}\r)
+{\cal O}\l(\f{p^2}{m_{KK}^2}\r)\,,
\eea
where $b_a$ are the 4D one-loop beta function coefficients
determined by the zero mode spectrum.
In ${\rm AdS}_5$ background,
$A^M_{++}$ gives a massless 4D vector,
$A^M_{--}$ a massless 4D real scalar, and $\psi_{zz}$ ($z=\pm$)
a massless 4D chiral spinor for any values of $k, R$ and $C$.
However  5D scalar field $\phi_{zz'}$
can give a zero mode for any value of $R$
only when $z=z'=+$ and its bulk and brane
masses satisfy
\bea
\label{scalarzeromode}
B_0=B_\pi\,,\quad
\sqrt{4+A^2}
=|\,2-B_0\,|\,.
\eea
Then $b_a$ are given by
\beq
\label{oneloopbeta}
b_a=-\f{11}{3}T_a(A^M_{++})+\f{1}{6}T_a(A^M_{--})
+\f{1}{6}T_a({\phi}^{(0)}_{++})+\f{2}{3}T_a(\psi_{++})
+\f{2}{3}T_a(\psi_{--}),
\eeq
where $\phi^{(0)}_{++}$ denotes
5D {\it real} scalar field having a zero mode.
Note that the conditions of (\ref{scalarzeromode})
are automatically satisfied
in supersymmetric theories as it should be.
In the following, we compute $\bar{\Delta}_a$ induced by generic 5D
scalar, Dirac fermion and vector fields with arbitrary
$Z_2\times Z^{\prime}_2$ boundary condition.

\begin{figure}
\centering
\epsfig{figure=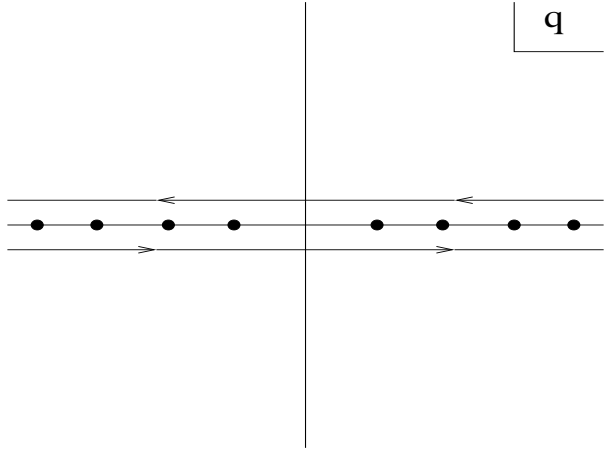,width=8cm,height=6cm}
\caption{\label{contour1} Contour $\leftrightharpoons$
in the complex $q$-plane. Bold dots represent the mass poles.}
\vspace{1cm}
\end{figure}

Regularizing a field theory on compact space
involves the regularization of the KK summation.
It is then convenient to convert the KK summation into an integral
 by introducing a pole function $P(q)$ \cite{GrootNibbelink:2001bx}
having the following properties:
(i) $P(q)$ has poles at $q = m_n$,
(ii) each pole has the residue 1,
(iii) there exists $\delta>0$ such that
$P\rightarrow B$ for $|{\rm Re}(q)|\rightarrow \infty$ and ${\rm Im}(q)
 > \delta$,
while $P\rightarrow -B$ for $|{\rm Re}(q)|\rightarrow \infty$ and
${\rm Im}(q)< -\delta$, where  $B$ is an {\it imaginary} constant.
These conditions uniquely determine the pole function.
In our case, it is given by
\beq
P(q) = \f{N'(q)}{2N(q)}\,,
\eeq
for which
\beq
\label{summation}
\sum_{m_n} \int d^4 p \, f ( p , m_n )=
\int_{\leftrightharpoons} \f{dq}{2 \pi i} \int d^4 p \,P(q) f(p, q),
\eeq
where $\leftrightharpoons$ denotes the contour depicted in Fig.
\ref{contour1}.

To obtain the 1-loop effective action of gauge field zero mode,
one needs to compute
\bea
\tr \ln \l( - D^2 + M^2(\Phi) +{F}_{\mu\nu} J^{\mu\nu}_{j} \r)
\eea
which contains the following two-point amplitude:
\bea
&&\int_\leftrightharpoons \f{d q}{2\pi i }
\,P(q)\,
\int \f{d^4 p}{(2\pi)^4} A^a_\mu (-p) A^a_\nu (p) T_a(\Phi) \nonumber\\
&& \times \l[ d(j) \int \f{d^4 k}{(2\pi)^4}
\f{g^{\mu\nu} \l((p+k)^2+q^2 \r) - \f{1}{2} (p+2k)^\mu (p+2k)^\nu }
{\l( k^2 +q^2\r) \l( (p+k)^2 +q^2 \r)} \r. \nonumber \\
&& \quad\quad \l.
-2C(j)\l(p^2 g^{\mu\nu}-p^\mu p^\nu \r)
\int \f{d^4 k}{(2\pi)^4} \f{1}{\l( k^2 +q^2\r) \l( (p+k)^2 +q^2 \r)}\,\r]
\nonumber \\
&&\equiv i \int \f{d^4 p}{(2\pi)^4} \,{\cal G}_a(p) A^a_\mu (-p)
\l(p^2 g^{\mu\nu} - p^\mu p^\nu \r)A^a_\nu(p)\,,
\label{trlogd2}
\eea
where $d(j)=(1,4,4)$ and $C(j)=(0,1,2)$  for $j=(0,1/2,1)$.
For the computation of the above integral,
it is convenient to split the pole function into two parts:
\beq
\label{decomposition}
P(q) = \t{P}(q) + {P}_{\infty}(q)\,,
\eeq
where $\t{P}\to {\cal O}(q^{-2})$ at $|q|\to \infty$.
Then $P_{\infty}$ can be written as
\beq
P_{\infty}(q) = -\f{A}{q} -B\epsilon({\rm Im}(q) )\,,
\eeq
where $\epsilon(x)=x/|x|$ and $A$ and $iB$ are some real constants, which
gives
\beq
\t{P}(q) = \f{N'(q)}{2N(q)} + \f{A}{q} +B\epsilon( {\rm Im}(q) ).
\eeq
With the decomposition (\ref{decomposition}),
all UV divergences appear in
the contribution from $P_{\infty}$ in a manner allowing simple
dimensional regularization,
while the contribution from $\t{P}$ is {\it finite}.

The 4D momentum integral $d^4p$
in (\ref{trlogd2}) exhibits
a branch cut on the imaginary axis of $q$. For the
contribution from $\t{P}$,
one can change the contour as in Fig. \ref{contour2}
since the contribution from the infinite half-circle vanishes.
After integrating by part,  we find that
the part of ${\cal G}_a$ from $\t{P}$ is given by
\bea
\Delta {\cal G}_a&=&
\f{T_a(\Phi)}{8\pi^2} \l( \f{1}{6} d(j) - 2 C(j) \r)
{\cal F}(q)\Big|_{q\to i\infty}
\nonumber \\
&&-\f{1}{8\pi^2} \int_0^1 dx \l( \f{1}{2} d(j) (1-2x)^2 - 2 C(j) \r)
{\cal F}(q)\Big|_{q = i\sqrt{x(1-x) p^2}}\,,
\eea
where
$$
{\cal F}(q)=\f{1}{2} \ln N + A \ln (-i q) + B q \,.
$$
The contribution from $P_{\infty}$ includes the log divergence
from the pole term $1/q$. This can be regulated by the standard
dimensional regularization of 4D momentum integral,
$d^4p\to d^Dp$, yielding
a $1/(D-4)$ pole. On the other hand, the step-function contribution
from $\epsilon({\rm Im}(q))$ involves a 5D momentum integral
which is linearly divergent, but it simply gives a finite result in
dimensional regularization.
Adding the divergent contribution from ${P}_{\infty}$ to the finite part
from $\t{P}$, we obtain
\bea
\label{twopoint}
{\cal G}_a&=& \f{T_a(\Phi)}{8\pi^2}\l[ \l( \f{1}{6} d(j) - 2 C(j) \r)
{\cal F}(q)\Big|_{q\to i\infty} \r. \nonumber\\
&&+\int_0^1 dx \, \l(-\f{1}{2}d(j)(1-2x)^2+2C(j)\r)
 \l(\f{1}{2} \ln N  \r) \Big|_{q = i\sqrt{x(1-x) p^2}}  \nonumber \\
&& \l.
+A \int_0^1 dx \l( -\f{1}{2} d(j) (1-2x)^2 +2C(j) \r) \l( \f{1}{D-4} \r)\r]\,.
\eea
In fact,  the values of $A$ and  ${\cal F}(q)$ at $q\to i\infty$
depend only on the $Z_2\times Z_2^{\prime}$ parity of the
corresponding 5D field, {\it not} on the spin of the field.
We then find
$$A= \l(-1/2, \, 0,\, 0,\,1/2\r)$$
for $Z_2\times Z_2^{\prime}$ parity $(Z_{\Phi},Z^{\prime}_{\Phi})=
(++,+-,-+,--)$
and
$$
{\cal F}\Big|_{q\to i\infty}=
\l(\,\f{1}{4}\pi kR-\f{1}{2}\ln k\,,\,
-\f{1}{4}\pi kR\,,\,\, \f{1}{4}\pi kR \,\,,\,
-\f{1}{4}\pi kR+\f{1}{2}\ln k\,\r)
$$
for the same $Z_2\times Z^{\prime}_2$ parity.

In order to get a physical result from (\ref{twopoint}), we still need to subtract
the $1/(D-4)$ pole.
When written in the position space of  5-th dimension,
$1/(D-4)$ term in (\ref{twopoint}) eventually leads to
a term $\propto (\lambda_0\delta(y)+\lambda_\pi\delta(y-\pi))F^a_{\mu\nu}
F^{a\mu\nu}/(D-4)$ in the 1-loop effective action.
(See Eqs. (\ref{divergence}) and (\ref{logdivergence})
for the definition of $\lambda_0$ and $\lambda_\pi$.)
Then the subtraction procedure should take into account that
the cutoff scales at $y=0$ and $\pi$ differ by the warp factor
$e^{-\pi kR}$. The correct subtraction scheme is to add a counter term
\bea
\label{sub}
\int d^4xdy\sqrt{G}\,
\f{1}{32\pi^2}&&
\l[\,
\lambda_0\l(\,\frac{1}{(D-4)}-\ln(\Lm)\,\r)
\f{\delta(y)}{\sqrt{G_{55}}}\r.
\nonumber \\
&&\l.+\lambda_\pi\l(\,\f{1}{(D-4)}-\ln(\Lm e^{-\pi kR})\,\r)
\f{\delta(y-\pi)}{\sqrt{G_{55}}}\,\r]
F^a_{\mu\nu}F^{a\mu\nu}\,,
\eea
which gives an extra $R$-dependent contribution $\propto
\lambda_\pi \pi kR$ to the low energy gauge coupling.
%In the previous calculations using dimensional regularization in ${\rm AdS}_5$,
%this piece has been overlooked,
This can be considered
in principle as a different choice of the bare
IR brane coupling $g^2_{\pi a}(\Lambda)$.
However if the 5D orbifold field theory is
regulated in $R$-independent manner, which is the most natural choice in
view of that $R$ is a dynamical field in 5D theory,
this extra piece should be considered as a part of calculable correction.
Also the strong coupling assumption
on the bare brane couplings\cite{Chacko:1999hg}, $g^2_{0a}(\Lambda)\approx
g^2_{\pi a}(\Lm)={\cal O}(8\pi^2)$, applies for the $R$-independent
part.
As we will see in the next section,
our subtraction scheme correctly reproduces
the results in supersymmetric case which can be obtained
by a completely independent method based on 4D effective SUGRA
whose regulator mass is $R$-independent.
We also explicitly show in Appendix B that
our subtraction scheme gives the precisely
same result as the $R$-independent Pauli-Villars
regularization for the case of 5D scalar QED.

With the prescription to compute the regularized
one-loop gauge coupling which has been discussed so far,
it is now straightforward
to compute $\Delta_a$ induced by generic 5D fields
with arbitrary $Z_2\times Z_2^{\prime}$ boundary condition.
The correction  due to 5D scalar fields is given by
\bea
\bar{\Delta}_a(\phi)=\f{1}{12}&&\l[
T_a(\phi_{++})\l\{\ln \l(\f{\Lm}{k}\r)
-3\int_0^1 du F(u)\ln N_{\phi_{++}}\l(\f{iu}{2} \sqrt{p^2}\r)
\r\}
\r.
\nonumber \\
&&-3T_a (\phi_{+-})
\int_0^1 du\, F(u)\ln N_{\phi_{+-}} \l(\f{iu}{2} \sqrt{p^2}\r)
\nonumber \\
&&-3T_a (\phi_{-+})
\int_0^1 du\, F(u)\ln N_{\phi_{-+}} \l(\f{iu}{2} \sqrt{p^2}\r)
\nonumber \\
&&\l.-T_a (\phi_{--})  \l\{ \ln \l(\f{\Lm}{k}\r)
+3 \int_0^1 du\, F(u)\ln N_{\phi_{--}} \l(\f{iu}{2} \sqrt{p^2}\r)
\r\}\,\r]
\eea
where the part with coefficient $T_a(\phi_{zz^{\prime}})$
represents the contribution from the loops of 5D scalar field
$\phi_{zz^{\prime}}$ and
$$
F(u)=u(1-u^2)^{1/2}.
$$
Here $N_{\phi_{zz^{\prime}}}$ ($z=\pm,z^{\prime}=\pm$)
 are the $N$-functions of
Eqs. (\ref{++bc}) and (\ref{otherbc}) for
$$(Z_\Phi,Z^{\prime}_\Phi,
s,r_0,r_\pi,\a)=(z,z^{\prime},4,B_0,B_\pi,\sqrt{4+A^2}).$$
The 1-loop corrections due to 5D fermion and vector fields are similarly
obtained to be
\bea
\bar{\Delta}_a(\psi)=\f{1}{3}&&\l[
T_a (\psi_{++})\l\{2\ln\l(\f{k}{p}\r)-\pi kR
+3\int_0^1 du \,G(u)
\ln N_{\psi_{++}} \l(\f{iu}{2} \sqrt{p^2}\r)
\r\}\r.\nonumber \\
&&+T_a (\psi_{+-})\l\{ -\pi k R
+3\int_0^1 du \,G(u)
\ln N_{\psi_{+-}} \l(\f{iu}{2} \sqrt{p^2}\r)
\r\}\nonumber \\
&&+T_a (\psi_{-+}) \l\{ \pi k R
+3\int_0^1 du \,G(u)
\ln N_{\psi_{-+}} \l(\f{iu}{2} \sqrt{p^2}\r)
\r\}
\nonumber \\
&&\l.+T_a (\psi_{--})\l\{2\ln\l(\f{k}{p}\r)-\pi kR
+3\int_0^1 du \, G(u)\ln N_{\psi_{--}} \l(\f{iu}{2} \sqrt{p^2}\r)
\r\}\,\r]\,,
\eea
\bea
\bar{\Delta}_a(A)=
\f{1}{12}&&\l[
T_a(A_{++})\l\{23\ln\l(\f{p}{\Lm}\r)+21\ln \l(\f{p}{k}\r)
+22\pi kR
+\int_0^1 du \,K(u)\ln N_{A_{++}}\l(\f{iu}{2} \sqrt{p^2}\r)
\r\}\r.\nonumber \\
&&+T_a(A_{+-})\l\{ -\pi k R  +\int_0^1 du\, K(u)
\ln N_{A_{+-}}\l(\f{iu}{2} \sqrt{p^2}\r)
\r\}\nonumber \\
&&+T_a(A_{-+}) \l\{\pi kR
+\int_0^1 du\, K(u) \ln N_{A_{-+}}\l(\f{iu}{2} \sqrt{p^2}\r)
\r\}\nonumber \\
&&\l.+T_a(A_{--})\l\{ 23 \ln \l(\f{\Lm}{k}\r)+2\ln\l(
\f{k}{p}\r)
-\pi kR
+\int_0^1 du\, K(u)
\ln N_{A_{--}}\l(\f{iu}{2} \sqrt{p^2}\r)
\r\}\,\r]\,,
\nonumber
\eea
where
$$
G(u)=u(1-u^2)^{1/2}-u(1-u^2)^{-1/2}\,,
$$
$$
K(u)=-9u(1-u^2)^{1/2}+24u(1-u^2)^{-1/2}.
$$
Here $N_{\psi_{++}},N_{\psi_{+-}},N_{\psi_{-+}}$ and $N_{\psi_{--}}$
are the $N$-functions of Eqs. (\ref{++bc}) and (\ref{otherbc})
for
\bea
(Z_\Phi,Z^{\prime}_\Phi,s,r,\a)&=&(-,-,1,C,|C-1/2|),\,
(+,-,1,-C,|C+1/2|),
\nonumber \\
&&(-,+,1,-C,|C+1/2|),\,(-,-,1,-C,|C+1/2|),
\nonumber
\eea
where $r=r_0=r_\pi$, and
$N_{A_{++}},N_{A_{+-}},N_{A_{-+}}$ and $N_{A_{--}}$
are the $N$-functions for
\bea
(Z_\Phi,Z^{\prime}_\Phi,s, r,\a)&=&(-,-,4,2,0),\,
(+,-,2,0,1),
\nonumber \\
&&(-,+,2,0,1),\,(-,-,2,0,1).
\nonumber
\eea
Note that $N_{\psi_{++}}$ and $N_{A_{++}}$ are given by
$N_{--}$ in Eq. (\ref{otherbc}), not $N_{++}$ in Eq. (\ref{++bc}).

For a practical application of the above results,
one  may consider the low momentum limit $p\ll m_{KK}$
where $m_{KK}$ denotes the {\it lowest} KK mass
which can be determined by the
corresponding $N$-function.
In such limit, one-loop gauge couplings can be written as
\bea
\f{1}{g_a(p)^2}&=&
\f{\pi R}{\hat{g}_{5a}^2}
+\f{1}{8\pi^2}\l[\,
\bar{\Delta}_a(p,A,B_0,B_\pi,C,k,R,\ln\Lambda)+{\cal O}(1)\,\r]
\nonumber \\
&=&\f{\pi R}{\hat{g}_{5a}^2}
+\f{1}{8\pi^2}\l[\Delta_a(A,B_0,B_\pi,C,k,R,\ln\Lambda)
+b_a\ln\l(\Lambda/p\r)
+{\cal O}\l(\f{p^2}{m_{KK}^2}\r)\r].
\eea
The results on $\Delta_a$ are summarized in Table I.
We also provide in Table II  the expressions of $\Delta_a$
induced by a scalar field with particular values of bulk and brane mass
parameters,
i.e. $B_0=B_\pi$ and
$\sqrt{4+A^2}=|2-B_0|$,
which corresponds to the scalar field in supersymmetric theory.

\section{4d supergravity calculation}

In \cite{Choi:2002wx,Choi:2002zi}, 1-loop low energy gauge couplings
 in ${\rm AdS}_5$  have been obtained in supersymmetric case using
the gauged $U(1)_R$ symmetry and chiral anomaly\cite{Arkani-Hamed:2001is} in
5D SUGRA in  ${\rm AdS}_5$\cite{Altendorfer:2000rr,Gherghetta:2000qt} and also
the known properties of
gauge couplings in 4D effective
SUGRA\cite{Kaplunovsky:1994fg}.
In this section, we confirm that  the results of
the previous section correctly reproduce the
SUGRA results when applied in supersymmetric case.

To proceed, let us briefly discuss supersymmetric 5D theory
on ${\rm AdS}_5$.
The theory contains two types of 5D supermultiplets other than
the SUGRA multiplet, one is the hypermultiplet
$H$ containing two 5D complex scalar fields $h^i$ ($i=1,2$) and
a Dirac fermion $\psi$, and the other is the vector
multiplet $V$ containing a 5D vector $A_M$, real scalar $\Sigma$ and
a symplectic Majorana fermion $\lambda^i$.
In supersymmetric model, all scalar fields have $B_0=B_\pi\equiv
B$
and $\sqrt{4+A^2}=|2-B|$ (see Eqs. (\ref{mass})
for the definitions of $B_{0,\pi}$ and
$A$) and their superpartner fermion has a kink mass parameter
$C=\pm (3-2B)/2$.
Also the $U(1)_R$ symmetry is gauged with the graviphoton $B_M$
in the following way:
\bea
&& D_Mh^i =\partial_M h^i-i\left(\frac{3}{2}(\sigma_3)^i_j-C\delta^i_j\right)
k\epsilon(y)B_M h^j +...\nonumber \\
&& D_M\psi=\partial_M \psi +iCk\epsilon(y)B_M\psi+...
\nonumber \\
&& D_M\lambda^{i}=\partial_M\lambda^{i}-i\frac{3}{2}(\sigma_3)^i_j k
\epsilon(y)B_M\lambda^{j}+...\,,
\eea
where $\psi$ has a kink mass $Ck\epsilon(y)$ and
the ellipses stand for the couplings with other gauge fields.
Taking into account the $Z_2\times Z^{\prime}_2$ parity, the
supermultiplet structure is given by
\bea
&&H_{zz^{\prime}}(C)=\l(\,h^1_{zz^{\prime}}(B=\f{3}{2}-C)\,,\,
 h^2_{\tilde{z}\tilde{z}^{\prime}}(B=\f{3}{2}+C)\,,\,
\psi^{\rm Dirac}_{zz^{\prime}}(C)\,\r)\,,\nonumber \\
&&V_{zz^{\prime}}=\l(\,A^M_{zz'}=(A^\mu_{zz^{\prime}},
A^5_{\t{z}\t{z}^{\prime}}(B=2))\,,\,
\lambda^i=\lambda^{\rm Dirac}_{zz^{\prime}}(C=\f{1}{2})\,,\,\Sigma_{\t{z}\t{z}^{\prime}}
(B=2)\,\r)\,,
\eea
where the subscripts $z,z^{\prime}$
denote the $Z_2\times Z^{\prime}_2$ parity,
$\t{z}=-z$, $\t{z}^{\prime}=-z^{\prime}$, $B$ is the brane mass parameter
and $C$ is the kink mass parameter.

Let us assume that our 5D theory is compactified
in a manner preserving $D=4$ $N=1$ supersymmetry.
This allows the low energy physics to be described by
4D effective SUGRA whose action can be written as
\beq \label{4daction} S_{4D}=\int d^4x \, \left[\,\int
d^4\theta  \, \left\{-3\exp \left(-\frac{K}{3}\right)\right\}
+\left( \int d^2\theta \, \frac{1}{4}
f_{a}\,W^{a\alpha}W^a_{\alpha}+h.c.\,\right)\,\right]\,, \eeq
where $W^a_{\alpha}$ is the chiral spinor superfield for the
4D gauge multiplet and we set the 4D gravity multiplet by their
vacuum values. The K\"ahler potential $K$ can be expanded in powers of
generic gauge-charged chiral superfield $Q$: \beq K\,=\,
K_0(\T,\T^*)+Z_{Q}(\T,\T^*)Q^*e^{-V}Q+...\,, \eeq where $\T$
denotes the radion superfield whose scalar component
is given by
$$
\T=R+iB_5\,,
$$
and the gauge kinetic function $f_a$ is  a
{\it holomorphic} function of $\T$.
Then  the 1-loop gauge couplings in effective 4D SUGRA
can be determined by
$f_a$ containing the 1-loop
threshold correction from massive KK modes and also
the tree-level K\"ahler potential $K$ \cite{Kaplunovsky:1994fg}:
\bea
\label{4dsugracoupling} \frac{1}{g^2_a(p)}\,&=&\,
\mbox{Re}(f_a)+\frac{b_a}{16\pi^2}
\ln\left(\frac{M_{Pl}^2}{e^{-K_0/3}p^2}\right) \nonumber \\
&&-\sum_{Q}\frac{T_a(Q)}{8\pi^2}\ln\left(e^{-K_0/3}Z_{Q}
\right) +\frac{T_a({\rm Adj})}{8\pi^2}\ln\left({\rm Re}(f_a)\right), \eea
where $b_a=\sum T_a(Q)-3T_a({\rm Adj})$ is the 1-loop beta
function coefficient and $M_{Pl}$ is the Planck scale of
$g_{\mu\nu}$ which defines
$p^2=-g^{\mu\nu}\partial_\mu\partial_\nu$.

Let us consider the 4D effective SUGRA of a 5D theory
which contains $H_{++},H_{+-},H_{-+},H_{--}$
as well as $V_{++},V_{+-},V_{-+},V_{--}$.
%\footnote{Here we limit
%the discussion to 5D theory without $V_{--}$ since $V_{--}$
%gives a massless  4D mode whose K\"ahler potential
%is not known at this moment.}.
The 5D vector multiplet
$V_{++}$ gives a massless 4D gauge multiplet
containing $A^\mu_{++}$ whose
low energy couplings are of interest for us, while
$V_{--}$ gives a massless
4D chiral multiplet containing $\Sigma_{++}+iA^5_{++}$.
$H_{++}$ and $H_{--}$ also give massless 4D chiral multiplets
containing $h^1_{++}$ and $h^2_{++}$, respectively,
whose tree level K\"ahler metrics
are required to compute the 1-loop gauge coupling (\ref{4dsugracoupling}).
Other 5D multiplets, i.e. $V_{+-},V_{-+},H_{+-}$ and
$H_{-+}$ do not give any massless 4D mode.
Let $Y_Q=e^{-K_0/3}Z_Q$ where
$Z_Q$ ($Q=H_{++},H_{--},V_{--}$) denote the
K\"ahler metric of the 4D massless chiral superfields coming from
the 5D multiplets $H_{++}, H_{--}$ and $V_{--}$, respectively.
Following Refs. \cite{Choi:2002zi,Marti:2001iw}, it is straightforward
to find the {\it tree level} $Z_{H_{++}},Z_{H_{--}}$
and also $f_a$ containing {\it the 1-loop threshold corrections}
from massive KK modes:
\bea \label{kahler}
&&
M_{Pl}^2=e^{-K_0/3}M_5^2=\f{M_5^3}{k}(1-e^{-k\pi(\T+\T^*)})
\,,\nonumber\\
&& Y_{H_{++}}\,=\,
\frac{M_5}{(\frac{1}{2}-C_{++})k} (e^{(\frac{1}{2}-C_{++})\pi
k(\T+\T^*)}-1)\,, \nonumber \\
&& Y_{H_{--}}\,=\,\f{M_5}{(\f{1}{2}+C_{--})k}
(e^{(\f{1}{2}+C_{--})\pi k(\T+\T^*)}-1)\,,
\nonumber \\
%%%%%%%%%
&& Y_{V_{--}}\,=\,\f{k }{M_5}
\f{1}{e^{\pi k (\T+\T^*)} -1}\,,
\nonumber \\
%%%%%%%%
&& f_{a}\,=\,\frac{\pi \T}{\hat{g}^2_{5a}} +\frac{z^{\prime}}{8\pi^2}\left(
\frac{3}{2}\sum_{V_{zz^{\prime}}}T_a(V_{zz^{\prime}})-
\sum_{H_{zz^{\prime}}}C_{zz^{\prime}}
T_a(H_{zz^{\prime}})\right)
k\pi \T\,, \eea
where $M_5$ is the 5D Planck scale,
and $C_{zz^\prime}$ is the kink mass of $H_{zz^\prime}$.
As was noted in \cite{Choi:2002zi}, the KK threshold correction to $f_a$
can be entirely determined by the chiral anomaly w.r.t the
following $B_5$-dependent phase transformation:
\bea \label{phase} &&
\lambda^{ai} \rightarrow
\l(e^{3ik|y|B_5\sigma_3/2}\r)^i_j \lambda^{aj}\,, \quad
\psi \rightarrow e^{-iCk|y|B_5}\psi\,.
\eea
Using the above results, we find the one-loop
gauge couplings at low momentum limit
$p\ll m_{KK}$ :
\bea
\f{1}{g_a(p)^2}=\f{\pi R}{\hat{g}_{5a}^2}+\f{1}{8\pi^2}\l[\,
\l(\Delta_a\r)_{\rm SUSY}
+\l(b_a\r)_{\rm SUSY}\ln(\Lambda/p)\,\r]\,,
\eea
where
\bea
\label{4dbulkcoupling}
\l(\Delta_a\r)_{\rm SUSY}
&=&-T_a(H_{++})\left[\,\ln\left(\frac{\Lm}{k}\right)+C_{++} \pi kR
+\ln \left(\f{
e^{(1-2C_{++})\pi kR}-1}{1-2C_{++}}\right)\,\right]
\nonumber \\
&&+C_{+-}T_a(H_{+-})\pi kR
-C_{-+}T_a(H_{-+})\pi kR
\nonumber \\
&&-T_a(H_{--})\l[ \,\ln \l(\f{\Lm}{k}\r)-C_{--}\pi kR
+\ln\l(\f{e^{(1+2C_{--})\pi kR}-1}{1+2C_{--}}\r)\,\r]
\nonumber \\
&&+T_a(V_{++})\left[\,\ln\left(M_5 \pi R \right) +\f{3}{2}\pi k R \,\right]
\nonumber \\
&&-\f{3}{2}T_a(V_{+-})\pi kR
+\f{3}{2}T_a(V_{-+})\pi kR
\nonumber \\
&&+T_a (V_{--}) \l[ \,\ln \f{M_5}{k} +\f{1}{2} k \pi R +
\ln \l(\f{1- e^{-2\pi kR}}{2}\r) \r]
\eea
and
\bea
\l(b_a\r)_{\rm SUSY}=-3T_a(V_{++})
+T_a(V_{--})+T_a(H_{++})+
T_a(H_{--}).
\nonumber
\eea
A rough estimate of $m_{KK}$ yields
$$m_{KK}\,\sim\, ke^{-\pi kR}$$
 for the bulk fields other than $H_{+-}$ or
$H_{-+}$.
On the other hand, $H_{+-}$ has
$$m_{KK}\,\sim\, ke^{-(\frac{1}{2}+C_{+-})\pi kR}
\,\,(C_{+-}\geq
1/2) \quad {\rm or} \quad
ke^{-\pi kR}
\,\,(C_{+-}\leq 1/2),
$$
while
$H_{-+}$ has
$$
m_{KK} \,\sim \,ke^{-(\frac{1}{2}-C_{-+})\pi kR}
\,\,(C_{-+}\leq -1/2)
\quad {\rm or}
\quad ke^{-\pi kR}\,\,(C_{-+} \geq -1/2).
$$
The above result
(\ref{4dbulkcoupling})
 obtained by 4D SUGRA analysis perfectly agrees with
the result that one would obtain using the results of
Tables I and II when $M_5$ is replaced by $\Lm$.
This provides a nontrivial
check for the results obtained in the previous section and
assures that our results are truely scheme-independent.

\section{conclusion}

In this paper, we have calculated  the full 1-loop corrections
to the low energy coupling of bulk gauge boson in ${\rm AdS}_5$ induced by generic
5D scalar, fermion and vector fields with arbitrary
$Z_2\times Z_2^{\prime}$ orbifold boundary condition.
The used calculation scheme is the background field method
with dimensional regularization.
We noted that
the subtraction scale for the log divergence at the IR brane
($y=\pi$) should be taken to be $\Lambda e^{-\pi kR}$
where $\Lambda$ is the subtraction scale for the UV brane
($y=0$).
We also considered supersymmetric case
to assure that our results correctly reproduce the results obtained
by a completely independent method based
on 4D effective supergravity analysis.

\bigskip\bigskip
{\bf Acknowledgement:}
We thank H. D. Kim for useful discussions. This work is supported in part
by BK21 Core program of MOE, KRF Grant No. 2000-015-DP0080, KOSEF Sundo-Grant,
and KOSEF through CHEP of KNU.

\bigskip
{\bf Note added:}
While this work was in completion, we received \cite{Contino:2002kc,Goldberger:new}
discussing the 1-loop gauge coupling renormalization due to 5D scalar loops in
${\rm AdS}_5$ background and its interpretation in the context of AdS/CFT correspondence
and also \cite{Falkowski:new} discussing the 1-loop renormalization in the context of
deconstructed ${\rm AdS}_5$.

\pagebreak
%%%%%%%%%%%%%%%%%%%%%%%%%%%%%%%%%%%%%%%%%%%%%%%%%%%%%%%%%%%%%%%%%%%%%%%%%%%%
%%                       APPENDIX                                        %%%
%%%%%%%%%%%%%%%%%%%%%%%%%%%%%%%%%%%%%%%%%%%%%%%%%%%%%%%%%%%%%%%%%%%%%%%%%%%%

%%%%%%%%%%%%%%%%%%%%%%%%%%%%%%%%%%%%%%%%%%%%%%
%  APPENDIX A. PROPERTIES OF N FUNCTION      %
%%%%%%%%%%%%%%%%%%%%%%%%%%%%%%%%%%%%%%%%%%%%%%

\begin{center}
{\bf Appendix A.\, Some properties of the $N$-functions}
\end{center}

In this appendix, we present some properties of
the $N$-functions, $N_{zz^\prime}$ ($z,z^\prime =\pm$), given in
Eqs. (\ref{++bc}) and (\ref{otherbc}).
Using
$Y_\a (x)= (\cos \a\pi J_\a (x) - J_{-\a} (x))/{\sin \a\pi }$
and also the fact that $N_{zz^\prime}$ are antisymmetric
under the exchange of $J_\a$ and $Y_\a$,
one can rewrite the $N$-functions as
\bea
N_{++}(q)&=&\frac{1}{\sin \alpha\pi}\l[\left\{(\frac{s}{2}-r_0)J_\alpha
\left(\f{q}{k}\right)+\f{q}{k}J^{\prime}_\a\left(\f{q}{k}\r)\r\}
\l\{(\f{s}{2}-r_\pi)J_{-\a}\l(\f{q}{T}\r)+
\f{q}{T}J^{\prime}_{-\a}\l(\f{q}{T}\r)\r\}\r.
\nonumber \\
&&\l.-\l\{(\f{s}{2}-r_\pi)
J_{\a}\l(\f{q}{T}\r)+\f{q}{T}J^{\prime}_{\a}\l(\f{q}{T}\r)\r\}
\l\{(\f{s}{2}-r_0)J_{-\a}\l(\f{q}{k}\r)+
\f{q}{k}J^{\prime}_{-\a}\l(\f{q}{k}\r)\r\}\r]\,,
\nonumber \\
N_{+-}(q)&=&\f{1}{\sin\a\pi}\l[\,
\l\{(\f{s}{2}-r_0)J_\a\l(\f{q}{k}\r)+\f{q}{k}J^{\prime}_\a\l(
\f{q}{k}\r)\r\}J_{-\a}\l(\f{q}{T}\r)
\r.\nonumber \\
&&\l.-J_\a\l(\f{q}{T}\r)\l\{
(\f{s}{2}-r_0)J_{-\a}\l(\f{q}{k}\r)+
\f{q}{k}J^{\prime}_{-\a}\l(\f{q}{k}\r)\r\}\,\r]\,,
\nonumber \\
N_{-+}(q)&=& \f{1}{\sin\a \pi}\l[\,\l\{(\f{s}{2}-r_\pi)J_\a\l(\f{q}{T}\r)
+\f{q}{T}J^{\prime}_{\a}\l(\f{q}{T}\r)\r\}J_{-\a}\l(\f{q}{k}\r)\r.
\nonumber \\
&&\l.-J_\a\l(\f{q}{k}\r)\l\{(\f{s}{2}-r_\pi)J_{-\a}\l(\f{q}{T}\r)
+\f{q}{T}J^{\prime}_{-\a}\l(\f{q}{T}\r)\r\}\,\r]\,,
\nonumber \\
N_{--}(q)&=& -\f{1}{\sin\a\pi}\l[\,
J_\a\l(\f{q}{k}\r) J_{-\a} \l(\f{q}{T}\r) - J_\a\l(\f{q}{T}\r) J_{-\a}\l(\f{q}{k}\r)
\,\r]\,,
\eea
where $T=ke^{-\pi kR}$.
Then using
$J_\a (x) = x^\a f(x^2)$,
one can easily see that all $N$-functions are even-functions:
$$
N_{zz^\prime}(q)=N_{zz^\prime}(-q).
$$
We already know $N_{zz'}(q)$ is
analytic near $q = 0$, allowing an expansion around $q=0$:
\beq
N_{zz'} (q) = Q_{zz'} + \f{q^2}{k^2} R_{zz'} + \Order{q^4}\,,
\eeq
where
\bea
Q_{++} &=&
\f{1}{\pi\a}\l[\, -(\a+r_\pi-\f{s}{2})(\a-r_0+\f{s}{2})e^{-\a \pi kR}
+(\a +r_0-\f{s}{2})(\a-r_\pi+\f{s}{2})e^{\a \pi kR}\,\r],
\nonumber \\
Q_{+-} &=&
\f{1}{\pi\a}\l[\, (\a-r_0+\f{s}{2})e^{-\a \pi kR}+
(\a+r_0-\f{s}{2})e^{\a \pi kR}\,\r],
\nonumber \\
Q_{-+} &=&
\f{1}{\pi\a}\l[\, (\a-r_\pi+\f{s}{2})e^{\a\pi kR}+
(\a+r_\pi-\f{s}{2})e^{-\a\pi kR}\,\r],
\nonumber \\
Q_{--} &=&
\f{1}{\pi \a} \l[ e^{\a \pi k R} - e^{-\a \pi k R} \r],
\nonumber \\
R_{++} &=&
-\f{1}{4\pi}\l[\,\f{1}{\a(\a-1)}\l\{
(2-\a-r_0+\f{s}{2})(\a-r_\pi+\f{s}{2})e^{\a \pi kR}\r.\r.
\nonumber \\
&& \quad\quad\quad\quad\quad\quad\quad\quad
+\l.(-2+\a+r_\pi-\f{s}{2})(\a-r_0+\f{s}{2})e^{(2-\a)\pi kR}\r\}
\nonumber \\
&&\quad\quad
+\f{1}{\a(\a+1)}\l\{(-\a-r_\pi+\f{s}{2})(2+\a-r_0+\f{s}{2})e^{-\a \pi kR}
\r.
\nonumber \\
&& \quad\quad\quad\quad\quad\quad\quad\quad
\l.\l.+(\a+r_0-\f{s}{2})(2+\a-r_\pi+\f{s}{2})e^{(\a+2)\pi kR}\r\}\,\r] ,
\nonumber \\
R_{+-} &=&
\f{1}{4\pi}\l[\,+\f{1}{\a(\a-1)}\l\{
(-2+\a+r_0-\f{s}{2})e^{\a\pi kR}+
(\a-r_0+\f{s}{2})e^{(2-\a)\pi kR}\r\}\r.
\nonumber \\
&&\l.-\f{1}{\a(\a+1)}\l\{(2+\a-r_0+\f{s}{2})e^{-\a \pi kR}
+(\a+r_0-\f{s}{2})e^{(2+\a)\pi kR}\r\}\,\r],
\nonumber\\
R_{-+} &=&
-\f{1}{4\pi}\l[\, \f{1}{\a(1-\a)}\l\{(-2+\a+r_\pi-\f{s}{2})
e^{(2-\a)\pi kR}+(\a-r_\pi+\f{s}{2})e^{\a\pi kR}\r\}\r.
\nonumber \\
&&\l.
+\f{1}{\a(1+\a)}\l\{(2+\a-r_\pi+\f{s}{2})e^{(2+\a)\pi kR}
+(\a+r_\pi-\f{s}{2})e^{-\a \pi kR}\r\}\,\r], \nonumber \\
R_{--} &=&
\f{1}{4\pi} \l[\,-\f{1}{\a(\a-1)} \l\{
e^{-(\a-2)\pi kR}-e^{\a\pi kR}\r\}
+\f{1}{\a(\a+1)}\l\{ e^{-\a\pi k R} - e^{(\a+2) \pi k R} \r\}\r].
\eea

The KK mass eigenvalue $m_n$ is determined
by the zeros of $N$-function: $N(m_n)=0$.
Obviously a 5D field has a massless 4D mode
iff $Q_{zz'} = 0$.
Generically, a nonzero KK mass eigenvalue starts to appear
from $m_n={\cal O}(T)$.
However in some special case, there can be nonzero mass
eigenvalues much smaller than $T=ke^{-\pi kR}$.
For instance, if $\a =\f{s}{2}-r_0$ and
$\a$ has  a large value, $Q_{+-}\sim e^{-\a \pi kR}$
and $R_{+-}\sim e^{\a \pi kR}$, giving a very light
state of $\Phi_{+-}$ with $m_n\sim ke^{-\a \pi kR}$.
Similarly, if $\a = r_\pi - \f{s}{2}$, $\Phi_{-+}$ can also have
a very small $m_n$.
However $\Phi_{--}$
does have neither a massless state
nor a very light state with $m_n\ll ke^{-\pi kR}$.

The asymptotic behavior of $N$-function at
$|q|\to \infty$ is essential for regularizing
the 1-loop gauge coupling.
Using the asymptotic formulae of Bessel functions:
\bea
J_\a (x) &\longrightarrow& \sqrt{\f{2}{\pi x}} \cos\l[ x - \l(\a+\f{1}{2}\r) \r] ,
\nonumber\\
Y_\a (x) &\longrightarrow& \sqrt{\f{2}{\pi x}} \sin\l[ x - \l(\a+\f{1}{2}\r) \r] ,
\nonumber
\eea
we find
\bea
N_{++}(q) \,\longrightarrow\,&&
\f{2qe^{\pi kR/2}}{\pi k}\sin \l(\f{(1-e^{\pi kR})q}{k}\r) ,
\nonumber \\
N_{+-}(q)\,\longrightarrow \, &&
\f{2}{\pi}e^{-\pi kR/2}\cos\l(\f{(1-e^{\pi kR})q}{k}\r) ,
\nonumber \\
N_{-+}(q) \,\longrightarrow\, &&
\f{2}{\pi}e^{\pi kR/2}\cos\l(\f{(1-e^{\pi kR})q}{k}\r),
\nonumber \\
N_{--}(q)\,\longrightarrow\, && -\f{2k}{\pi q}
e^{-\pi kR/2}\sin\l(\f{(1-e^{\pi kR})q}{k}\r)\,.
\nonumber
\eea

%%%%%%%%%%%%%%%%%%%%%%%%%%%%%%%%%%%%%%%%%%%%%
%   Appendix , Comparison with PV        %%%
%%%%%%%%%%%%%%%%%%%%%%%%%%%%%%%%%%%%%%%%%%%%%
\begin{center}
{\bf Appendix B. \, Comparison with Pauli-Villars Regularization}
\end{center}

A natural regularization in 5D theory is to cut off 5D momentum
in the 5D metric frame of $G_{MN}$: $-G^{MN} \p_M \p_N < \Lm^2$.
In AdS background, this would correspond
to an effective $y$-dependent cut-off of 4D momentum in the 4D metric
frame of $g_{\mu\nu}$: $p^2=-g^{\mu\nu}\p_\mu\p_\nu < e^{-2kR|y|}\Lm^2$.
In dimensional regularization, such feature is not
manifest, but can be taken into account by choosing
the subtraction scale $\sim \Lm e^{- k R \t{y}}$ where $\t{y}=0$ or $\pi$
is the location of log divergence.
On the other hand, such feature is rather manifest
in Pauli-Villars (PV) regularization in which $\Lambda$
corresponds to a 5D regulator mass.
In this appendix, we compare our result using the dimensional
regularization with the subtraction scheme (\ref{sub})
to the PV result for scalar QED.
For simplicity, we consider the massless
scalar QED with $Z_2\times Z_2^\prime$ parity $(++)$.

\begin{figure}
\centering
\epsfig{figure=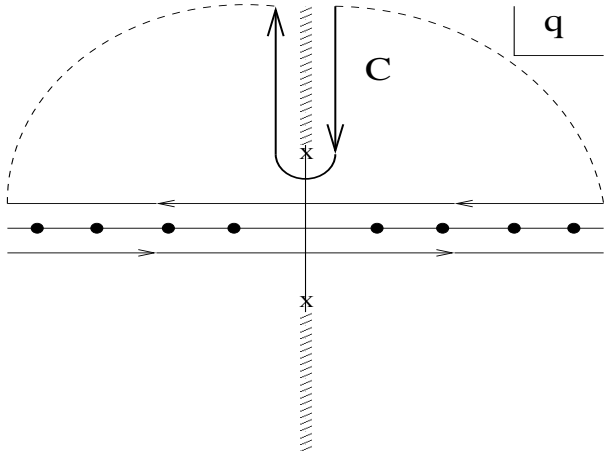,width=8cm,height=6cm}
\caption{\label{contour2} For the  contribution from $\tilde{P}(q)$,
the contour $\leftharpoonup$ can be deformed to the contour $C$ represented by
the bold line since the contribution vanishes
on the dotted infinite half circle.
Hatched lines on the imaginary axis are logarithmic branch-cuts.
After integrating by parts, the point {\bf x} where the branch-cut starts
 becomes a simple pole. Then the integral along $C$
is given by the values of integrand at the boundary of $C$ at infinity
and the residue value at the point {\bf x}. The integral along
$\rightharpoondown$ can be similarly treated in the lower half plane.  }
\end{figure}

In PV scheme, the UV divergence is regulated by a  PV
regulator with 5D mass $\Lambda$ which has
the same $Z_2\times Z_2^\prime$ boundary condition as
$\phi$ but opposite statistics:
\beq
\sum_n \int \f{d^4 p}{(2\pi)^4} f( p, m_n ) \longrightarrow
\sum_n \l\{ \int \f{d^4 p}{(2\pi)^4} f(p,m_n)-\int \f{d^4 p}{(2\pi)^4} f(p,M_n)
\r\},
\eeq
where $M_n$ is the KK spectrum for the PV regulator.
We convert the summation into an integral using the pole functions:
$$P_\phi = \f{N'_\phi}{2N_\phi}
\,,\quad
P_{\rm PV}=\f{N'_{\rm PV}}{2N_{\rm PV}}\,,
$$
and then the regulated amplitude is given by
\beq
\int_\leftrightharpoons \f{d q}{2\pi i} P_{\rm reg} (q) \int
\f{d^4 p}{(2\pi)^4}f(p, q),
\eeq
where $P_{\rm reg}(q) \equiv P_\phi (q) - P_{\rm PV} (q)$.
Since $N_\phi$ and $N_{\rm PV}$ are the same limiting behavior
at $|q|\to \infty$, $P_{\rm reg}(q)$ vanishes at infinity.
After a partial integration along $q$, we find
\bea
\bar{\Delta}_{\rm PV} &=&
8\pi^2 \int_C \f{d q}{2 \pi i}
 \l(
 \f{1}{2} \ln N_\phi - \f{1}{2}\ln N_{\rm PV} \r)
\nonumber \\
&& \times \f{d}{dq}
\l\{ \f{1}{2} \int dx (1-2x)^2 \f{1}{(4\pi)^2} \ln \l( x(1-x) p^2 + q^2 \r)
\r\} \nonumber \\
&=&
 -\f{1}{4} \int dx (1-2x)^2 (\ln N_\phi - \ln N_{\rm PV})\Big|_{q=i
\sqrt{x(1-x)p^2}},
\eea
where $C$ is the contour line described in Fig. \ref{contour2}.
For $q \ll ke^{-\pi kR}$,
\bea
N_\phi \, &\approx &\, \f{q^2}{k^2} e^{\pi k R}
\l(\f{e^{\pi k R} - e^{-\pi k R}}{\pi}\r),\\
N_{\rm PV} \,&\approx&\, \f{(\a-2)(\a+2)}{\pi \a} \l( e^{-\a \pi k R}
- e^{\a\pi k R}\r),
\eea
where $\a=\sqrt{4+\Lm^2/k^2}$.
For $\Lm \gg k$, $\a\approx \Lm/k$, so
\beq
\ln N_{\rm PV} \approx \Lambda \pi R + \ln(\Lm/k)\,.
\eeq
After subtracting the power-law divergent part
which is regularization scheme dependent,
we find
\beq
\Delta_{\rm PV}\equiv \bar{\Delta}_{\rm PV}
-b_a\ln(\Lambda/p)
= -\f{1}{12} \l[\, \ln (\Lm/k)
+\pi k R
+\ln\l(\f{e^{\pi k R} - e^{-\pi k R}}{2\pi}\r)\,\r],
\eeq
which is precisely same as the result in Table II
for a massless real $\phi_{++}$ with $A=B_0=B_\pi=0$.
In scalar QED, a charged scalar field should be complex,
so gives a loop correction twice of the above result.

%%%%%%%%%%%%%%%%%%%%%%%%%%%%%%%%%%%%%%%%%%%%%%%%%%%%%%%%%%%%%%%%%%%%%%%%
%%                    p^2 << 1st KK mass                              %%
%%%%%%%%%%%%%%%%%%%%%%%%%%%%%%%%%%%%%%%%%%%%%%%%%%%%%%%%%%%%%%%%%%%%%%%%

\begin{table}
\caption{\label{lowenergylimit} One loop corrections for $p\ll m_{KK}$
where $m_{KK}$ is the lowest nonzero KK mass. One-loop gauge couplings
are given by $\f{1}{g_a^2}=\f{\pi R}{\hat{g}_{5a}^2}+
\f{1}{8\pi^2}\l[\,
\Delta_a+b_a\ln(\Lambda/p)\,\r]$ where
$b_a$ is the 4D one-loop beta function coefficient
due to zero modes (see (\ref{oneloopbeta})).
Here $\phi_{zz'}$ ($z,z'=\pm$) stand for real scalar fields
which do not have zero mode since they have generic mass-squares given by
$A^2k^2+\f{2k}{R}\l[
B_0\delta(y)-B_\pi\delta(y-\pi)\r]$,
while
$\phi^{(0)}_{++}$ denotes a scalar field with zero mode, i.e.
a scalar field with $(++)$ parity, $B_0=B_\pi\equiv B$ and $
\sqrt{4+A^2}=
|2-B|$. Here $\a\equiv\sqrt{4+A^2}$.}
%\setlongtables

\bigskip

\begin{tabular}%longtable
{|c|c|l|}
Type & $(zz')$ & \quad
\quad\quad\quad\quad\quad $\Delta_a (A,B_0,B_\pi,C,k,R,\ln \Lm)$ \\
\hline
\hline
& &
\\
\,\, real \quad\quad & $(++)$ &
$-\f{1}{12}T_a(\phi^{(0)}_{++})
\l[\ln\l(\Lm/k\r)+\pi kR+\ln\l(\f{e^{(1-B)\pi kR}-e^{-(1-B)\pi kR}}
{2(1-B)}
\r) \r]$\\
scalar   &  &
\\
& &
$+\f{1}{12} T_a(\phi_{++}) \l[\ln(\Lm/k)
-\ln \l(\f{(\a+B_0-2)(\a-B_\pi+2)e^{\a \pi kR}
-(\a+B_\pi-2)(\a-B_0+2)e^{-\a \pi kR}}{2\a}\r) \r]
$
 \\
& &
\\
\cline{2-3}
& &
\\
 & $(+-)$ &
$-\f{1}{12} T_a(\phi_{+-})\ln\l(
\f{(\a+B_0-2)e^{\a \pi kR}+
(\a-B_0+2)e^{-\a\pi kR}}{2\a}\r)
$ \\
& &
\\
\cline{2-3}
& &
\\
 & $(-+)$ &
$-\f{1}{12} T_a(\phi_{-+}) \ln \l(
\f{(\a-B_\pi+2)e^{\a\pi kR}+
(\a+B_\pi-2)e^{-\a\pi kR}}{2\a}\r)
$ \\
& &
\\
\cline{2-3}
& & \\
& $(--)$ &
$
-\f{1}{12} T_a (\phi_{--}) \l[\ln (\Lm/k)
+\ln \l(\f{e^{\a \pi k R} - e^{-\a \pi k R}}{2\a}\r)
\r]
$
\\
& & \\
\hline
& & \\
Dirac & $(++)$&
$
-\f{2}{3} T_a (\psi_{++})  \l[ \,\ln(\Lm/k) +\f{1}{2}\pi kR
+\ln \l(\f{\displaystyle e^{\l(C-\f{1}{2}\r)\pi k R}
-e^{-\l(C-\f{1}{2}\r)\pi k R}}
{2\l(C-\f{1}{2}\r)}\r)\, \r]
$  \\
spinor & & \\
\cline{2-3}
& & \\
 & $(+-)$ &
$
\,\,\,\f{2}{3} T_a (\psi_{+-})  ~C\pi k R
$ \\
& & \\
\cline{2-3}
& & \\
& $(-+)$ &
$
-\f{2}{3} T_a (\psi_{-+}) ~C \pi k R
$\\
& & \\
\cline{2-3}
& & \\
& $(--)$ &
$
-\f{2}{3} T_a (\psi_{--})  \l[\,
\ln (\Lm/k)+\f{1}{2}\pi kR
+\ln \l(
\f{\displaystyle e^{\l(C+\f{1}{2}\r)\pi k R}
-e^{-\l(C+\f{1}{2}\r)\pi k R}}
{2
\l(C+\f{1}{2}\r)} \r)\,
\r]
$ \\
& &
\\
\hline
& & \\
vector  & $(++)$ &
$
\f{1}{12} T_a(A^M_{++}) \l[ \,21 \ln (\Lm \pi R)+22 \pi kR\,\r]
$ \\
& & \\
\cline{2-3}
& & \\
& $(+-)$ &
$
-\f{11}{6} T_a(A^M_{+-}) ~\pi k R
$ \\
& & \\
\cline{2-3}
& & \\
& $(-+)$ &
$
\,\,\,\f{11}{6} T_a(A^M_{-+}) ~\pi k R
$
\\
& & \\
\cline{2-3}
& & \\
& $(--)$ &
$
\f{1}{12} T_a(A^M_{--}) \l[ \,21 \ln (\Lm\pi R)
 - \pi k R
+ 21 \ln \l( \f{\displaystyle e^{\pi k R}- e^{-\pi kR}}{2\pi kR} \r)
\r]
$\\
& & \\
\end{tabular}%longtable}
\end{table}

\vskip 0.5cm
%%% TABLE 2 %%%%%%%%%%%
\begin{table}
\caption{\label{scalarspecialcase} 5D scalar contribution
for $p\ll m_{KK}$ when $B_0 = B_\pi\equiv B$ and $\a\equiv
\sqrt{4+A^2} = |2-B|$.}
\bigskip
\begin{tabular}{|c|l|}
& \\
\,\, $(++)$\,\, &
$-\f{1}{12} T_a (\phi_{++}) \l[\, \ln (\Lm/k) +\pi kR
+\ln \l(\f{\displaystyle e^{(1-B) \pi kR}-
e^{-(1-B) \pi k R}}{2(1-B)}\r)\, \r]$\\
& \\
\hline
& \\
\,\, $(+-)$ \,\, & \,\, \,$\f{1}{12} T_a (\phi_{+-}) (2-B) \pi k R $ \\
& \\
\hline
& \\
\,\, $(-+)$ \,\, & $-\f{1}{12} T_a (\phi_{-+}) (2-B) \pi k R $ \\
& \\
\hline
& \\
\,\, $(--)$ \,\, &
$-\f{1}{12} T_a(\phi_{--}) \l[ \,\ln (\Lm/k)
+\ln \l(\f{\displaystyle
e^{(2-B) \pi k R} - e^{-(2-B) \pi k R }}{2 (2-B)}\r)\,
\r]$
\\
&
\end{tabular}
\end{table}

%%%%%%%%%%%%%%%%%%%%%%%%%%%%%%%%%%%%%%%%%%%%%%%%%%%%
%%%                 FIGURES                      %%%
%%%%%%%%%%%%%%%%%%%%%%%%%%%%%%%%%%%%%%%%%%%%%%%%%%%%
% Figure1
% Figure2

%%%%%%%%%%%%%%%%%%%%%%%%%%%%%%%%%%%%%%%%%%%%%%%%%%%%%%%%%%%%%%%%%%%%%%%%%%%%%%

\end{document}